\documentclass[twocolumn,amsmath,amssymb]{revtex4}
\usepackage{graphicx}
\usepackage{dcolumn}
\usepackage{bm}
\begin{document}

\title{Interacting Dirac Fermions on a Topological Insulator in a Magnetic Field}
\author{Vadim M. Apalkov}
\affiliation{Department of Physics and Astronomy, Georgia State University,
Atlanta, Georgia 30303, USA}
\author{Tapash Chakraborty$^\ddag$}
\affiliation{Department of Physics and Astronomy,
University of Manitoba, Winnipeg, Canada R3T 2N2}

\date{\today}
\begin{abstract}
We have studied the fractional quantum Hall states on the surface of a 
topological insulator thin film in an external magnetic field, where 
the Dirac fermion nature of the charge carriers have been experimentally 
established only recently. Our studies indicate that the fractional 
quantum Hall states should indeed be observable in the surface Landau 
levels of a topological insulator. The strength of the effect will 
however be different, compared to that in graphene, due to the finite 
thickness of the topological insulator film and due to the admixture of 
Landau levels of the two surfaces of the film. At a small film thickness, 
that mixture results in a strongly non-monotonic dependence of the 
excitation gap on the film thickness. At a large enough thickness 
of the film, the excitation gap in the lowest two Landau 
levels are comparable in strength.
\end{abstract}
\maketitle

In recent years, there has been an upsurge of interest on the unusual 
magnetic properties of Dirac fermions in condensed matter systems. 
Graphene \cite{Geim_Nobel,abergeletal} is one such system where those 
properties were exhaustively investigated theoretically \cite{old_magnetic} 
and experimentally \cite{Jiang_07,Zhang_06}. The most remarkable of 
those are the existence of a Landau level at zero energy having no 
magnetic field dependence, and the square root dependence of higher energies 
on the Landau level (LL) index and the magnetic field \cite{Jiang_07,Zhang_06}. 
A direct consequence of these is the unusual half-integer quantum Hall 
effect (QHE). The role of Coulomb interacions between Dirac fermions in 
graphene, in the fractional QHE (FQHE) regime, was first studied by us
theoretically \cite{Apalkov_06}. The effect was later observed in 
the experiments \cite{Abanin_10,Ghahari_11}.

Linear dispersion relation in the band structure of massless Dirac fermions, 
first observed in monolayer graphene \cite{wallace,Geim_Nobel,abergeletal}, 
is also expected to be present in special insulators with topologically 
protected surface states \cite{hasan_2010,qi}. The surface states in these 
topological insulators (TIs) are gapless with linear relativistic dispersion 
relation, and therefore bear a close analogy, albeit with some major 
differences \cite{note_1}, to the energy spectra in graphene. In an external 
magnetic field, the energetics of the surface Landau levels in the TIs
is also similar to that of the Landau levels in graphene. Recently, a direct 
observation of Landau quantization in a TI, Bi$^{}_{1-x}$Sb$^{}_x$ thin film, 
was reported, which was found to have the same behavior as in graphene, 
but are found to be suppressed by surface impurities, thereby indicating its 
two-dimensional (2D) nature \cite{Topo_LL_1,Topo_LL_2}. Although the low-energy 
dynamics of the surface states is similar to that of graphene, there is an 
important difference between these two systems. While the electronic states in 
graphene are strictly two-dimensional and localized within a single graphene 
layer, the surface states of a TI have finite width 
in the growth direction. This modifies the inter-electron interaction 
potential and changes the properties of the FQHE states. In traditional 
non-relativistic electron systems, an increase in width of the 2D layer 
results in a reduction of the FQHE gaps, which in turn reduces the 
stability of the corresponding incompressible states \cite{FQHE_book}. 
Therefore, we should expect that the FQHE gaps in the TIs
are also influenced by the finite width, in contrast to the case in 
graphene. Experimental observation of the Landau quantized topological 
surface states, in particular the field-independent LL at the Dirac point 
\cite{Topo_LL_1,Topo_LL_2}, the signature effect of Dirac fermions, has 
paved the way for future observation of the QHE in the surface states of 
the TI. The nature of the FQH states in such a system is the subject of
this Letter.

To analyze the properties of the FQHE in the surface states of the TIs, we 
begin with the low-energy effective Hamiltonian already introduced 
in the literature \cite{liu_2010,zhang_2009}. The Hamiltonian has the 
matrix form of size $4\times 4$ and is given by
\begin{equation}
{\cal H}^{}_{\rm TI}=\epsilon(\vec{p})+ \left(  
\begin{array}{cc}
 M(\vec{p})\sigma^{}_z+A^{}_1 p^{}_z\sigma^{}_x & A^{}_2 p^{}_{-}\sigma^{}_x \\
A^{}_2 p^{}_{+}\sigma^{}_x & M(\vec{p})\sigma^{}_z - A^{}_1 p^{}_z\sigma^{}_x 
\end{array}
\right),
\label{HTI}
\end{equation}
where $\sigma^{}_i$ are the Pauli matrices, 
$\epsilon (\vec{p})=C^{}_1 + D^{}_1 p_z^2 + D^{}_2 (p_x^2 + p_y^2) $, 
$M(\vec{p})= M^{}_0 - B^{}_1 p_z^2 - B^{}_2 (p_x^2 + p_y^2)$. Here we consider
a prototypical TI, $\mbox{Bi}^{}_2\mbox{Se}^{}_3$, with one surface state 
containing a single isotropic Dirac cone at the center of the Brillouin zone. 
For this system the constants are \cite{zhang_2009}, 
$A^{}_1 = 2.2$ eV$\cdot$\AA,  $A^{}_2 = 4.1$ eV$\cdot$\AA,
$B^{}_1 = 10$ eV$\cdot$\AA$^2$, $B^{}_2 = 56.6$ eV$\cdot$\AA$^2$,
$C^{}_1 = -0.0068$ eV, 
$D^{}_1 = 1.3 $ eV$\cdot$\AA$^2$, 
$D^{}_2 = 19.6 $ eV$\cdot$\AA$^2$, and 
$M^{}_0 = 0.28$ eV. 
The topological insulator film has a finite thickness of $L^{}_z$, i.e.,
the two surfaces of the film are at $z=0$ and $z=L^{}_z$. The four-component 
wavefunctions corresponding to the Hamiltonian (\ref{HTI}) determine the 
amplitudes of the wavefunctions at the positions of Bi and Se atoms: 
$(\mbox{Bi}^{}_{\uparrow}, \mbox{Se}^{}_{\uparrow}, \mbox{Bi}^{}_{\downarrow}, 
\mbox{Se}^{}_{\downarrow})$, where the arrows indicate the direction of the 
electron spin. 

\begin{figure}
\begin{center}\includegraphics[width=7cm]{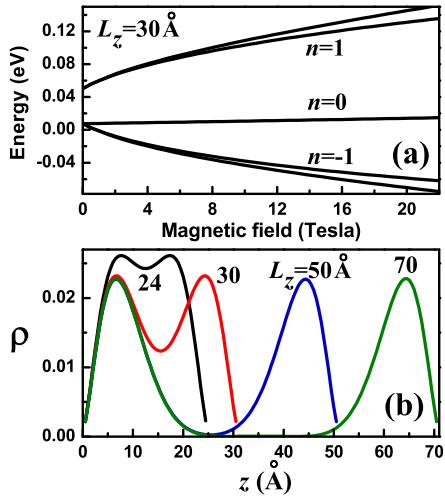}
\end{center}
\caption{\label{Topological0}
(a) The lowest surface Landau levels of a TI film shown for the film thickness 
$L_z = 30$ \AA. For each $n$ there are two LLs of the TI film, belonging to 
two surfaces of the film. (b) The electron density along the $z$ axis for 
one of the $n=1$ Landau levels and for different values of the thickness, 
$L_z$, of the $\mbox{Bi}^{}_2\mbox{Se}^{}_3$ film. The numbers next to the 
lines are the values of $L_z$. The magnetic field is at 15 Tesla. 
}
\end{figure}

The external magnetic field, $B$, along the $z$-direction produces surface 
Landau levels which can be found from the Hamiltonian matrix by replacing 
the $x$ and $y$ components of the momentum by the generalized momentum
\cite{yang_2011} and introducing the Zeeman energy, $\Delta^{}_z=\frac12 g_s
\mu^{}_B B$. Here $g_s\approx 8$ is the effective $g$-factor of surface 
states \cite{liu_2010,wang_2010} and $\mu^{}_B$ is the Bohr magneton. 
For the surface Landau level $n$ the wavefunctions are of the form 
\begin{equation}
\Psi^{\rm (TI)}_{n}  = 
\left( \begin{array}{c}
 \chi^{(1)}_n (z) \phi^{}_{|n|-1,m} \\
  \chi^{(2)}_n (z)   \phi ^{}_{|n|-1,m} \\  
  i \chi^{(3)}_n (z)  \phi^{}_{|n|,m} \\
  i \chi^{(4)}_n(z) \phi ^{}_{|n|,m}  
\end{array}  
 \right),
\label{fTI}
\end{equation} 
for $n>0$. For $n=0$, the functions $\chi^{(1)}_{n=0}$ and $\chi^{(2)}_{n=0}$ 
vanish identically. Here $\phi^{}_{n,m}$ are wavefunctions of conventional 
non-relativistic LLs with index $n$ and some other intra-Landau level index, 
$m$, for example, the $z$ component of the angular momentum. 
The functions $\chi^{(i)}_n(z)$ satisfy the following eigenvalue equation 
\begin{eqnarray}
& & \varepsilon \hat{\chi}^{}_n(z) =
\left(
\begin{array}{cc}
\epsilon_{z,n}^{(+)} + M^{}_{z,n} \sigma^{}_z & 0   \\
0  &
\epsilon_{z,n+1}^{(-)} + M^{}_{z,n+1} \sigma^{}_z
\end{array}
\right)  \hat{\chi}_n(z) \nonumber \\
& & + \left(
\begin{array}{cc}
i A^{}_1\sigma^{}_x \frac{d}{dz} & -\frac{\sqrt{2(n+1)}}{\ell^{}_0 } 
A^{}_2 \sigma^{}_x   \\
-\frac{\sqrt{2(n+1)}}{\ell^{}_0} A^{}_2 \sigma^{}_x  &
- i A^{}_1 \sigma^{}_x \frac{d}{dz}
\end{array}
\right)\hat{\chi}^{}_n(z)
\label{Tseq4},
\end{eqnarray}
where $\epsilon^{(\pm)}_{z,n}=C^{}_1 + D^{}_2 (2n+1)/\ell_0^2 -
D^{}_1 \frac{d^2}{dz^2} \pm \Delta^{}_z$ and $M^{}_{z,n}= M^{}_0- B^{}_2
(2n+1)/\ell_0^2 - B^{}_1 \frac{d^2}{dz^2}$. Here $\ell^{}_0=\sqrt{e\hbar/c B}$ is
the magnetic length and $\hat{\chi}^{}_n = (\chi^{(1)}_n,
\chi^{(2)}_n,\chi^{(3)}_n,\chi^{(4)}_n)^{T}$.
The solution of the system of equations (\ref{Tseq4})
determines the Landau energy spectrum of the system and
the width in the $z$-direction of the Landau level wavefunctions. This width 
depends on the Landau level index, the strength of the magnetic field, and 
the thickness of the TI film. Each surface of the TI film has a corresponding 
set of surface LLs. The LLs of two surfaces of TI are coupled, which is more 
pronounced for a narrow TI film, $L_z \lesssim 30$ \AA. For TI films of large 
thickness, the LLs of two surfaces can be considered as decoupled. In 
Fig.~\ref{Topological0} we illustrate the properties of the surface LLs, 
which were obtained from the numerical solution of 
Eq.~(\ref{Tseq4}). In what follows, we consider only the lowest surface 
LLs with indices $n=0$ and $n=1$. Strong FQHEs can be observed only in these
LLs. In Fig.~\ref{Topological0}(a), the LL energy spectrum is shown as a 
function of the magnetic field, $B$, for a TI film of thickness $L_z=30$ \AA. 
The $\sqrt{B}$-dependence of the energy of $n=1$ ($n=-1$) LLs is visible, 
which is a specific property of the relativistic dispersion law of the surface 
states of the TI. For each value of $n$ there are two degenerate surface LLs, 
belonging to the two surfaces. The degeneracy of these LLs is lifted due to 
their coupling. For $n=0$, the inter-Landau level coupling is relatively weak
and the $n=0$ LL remains almost doubly degenerate. The coupling of the surface 
LLs is illustrate in Fig.~\ref{Topological0}(b), where the electron density, 
$\rho (z) = |\chi^{(1)}_n|^2 + |\chi^{(2)}_n|^2 + |\chi^{(3)}_n|^2 
+|\chi^{(4)}_n|^2$, is shown for different thicknesses of the TI film and 
for one of the $n=1$ LLs. For a small thickness of the TI film the surface LLs 
strongly overlap in space, which results in a strong inter-LL coupling
and large electron density within the whole TI film. With increasing film
thickness the surface LLs become decoupled with strong localization at the 
surfaces of the TI film and zero electron density in the bulk region of the film. 

In the FQHE regime the electrons partially occupy a single LL and the properties 
of such a system are completely determined by the electron-electron interaction 
potential within the corresponding LL \cite{FQHE_book}. The interaction potential 
projected on a given LL is determined by the Haldane pseudopotentials 
\cite{Haldane_83}, which are the energies of two electrons with relative angular 
momentum $m$. With the known wavefunctions (\ref{fTI}) of the LL with index $n$ 
the pseudopotentials can be readily evaluated.

\begin{figure}
\begin{center}\includegraphics[width=7cm]{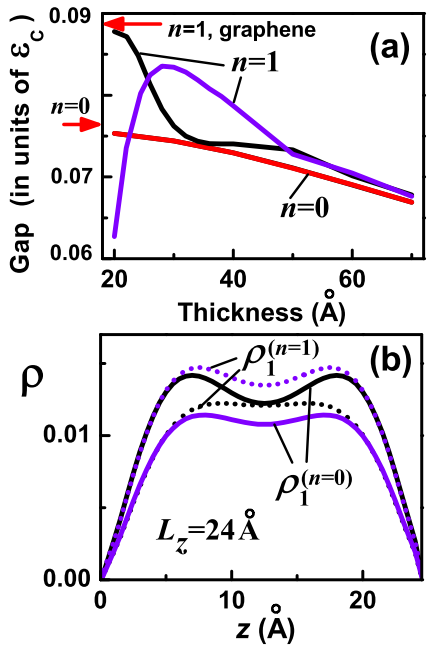}
\end{center}
\caption{\label{Topological2}
(a) The $\nu =\frac13$ FQHE gap, shown for different Landau levels of 
a TI film as a function of the film thickness. The magnetic field is 15 
Tesla. The excitation gaps were obtained numerically for a finite-size 
system of $N=9$ electrons and the parameter of the sphere $S=12$ 
\cite{Haldane_83}. The energy is shown in units of the Coulomb energy, 
$\varepsilon^{}_C = e^2/\kappa \ell^{}_0$. The red arrow indicates 
the FQHE gap in the $n=0$ Landau level of graphene. (b) The electron 
densities $\rho_{1}^{(n=1)}(z)$ and $\rho_{1}^{(n=0)}(z)$ of 
two $n=1$ Landau levels at a film thickness of $L_z=24$ \AA. The black 
(solid and dotted) lines correspond to the $n=1$ Landau level in panel 
(a) (black line), while the blue (solid and dotted) lines correspond to 
the $n=1$ Landau level in panel (a) (blue line). The densities 
$\rho_{1}^{(n=1)}(z)=|\chi^{(3)}_n(z)|^2 +|\chi^{(4)}_n(z)|^2$ and 
$\rho_{1}^{(n=0)}(z)=|\chi^{(1)}_n(z)|^2 +|\chi^{(2)}_n(z)|^2$ show the 
occupations of the $n=1$ and $n=0$ non-relativistic 
Landau level functions, respectively. 
}
\end{figure}

Just as in graphene, the $n=0$ Landau level in a TI is identical to the
$n=0$ traditional (non-relativistic) Landau level but with a finite width, 
while the $n>0$ LL in the TI is a mixture of the $n$ and $n-1$ non-relativistic 
states. In graphene, such a mixture results in the strongest FQHE being in the 
$n=1$ LL \cite{Apalkov_06}. Similar behaviour is also expected in a TI film. 
To study the properties of the FQHE in a TI we numerically evaluate the energy spectrum 
of a finite $N$-electron system in the spherical geometry \cite{Haldane_83}. 
The radius of the sphere is $\sqrt{S}\ell^{}_0$, where 
$2S$ is the number of magnetic fluxes through the sphere in units of the 
flux quantum. For a given number of electrons, the radius of the sphere 
determines the filling factor of the system. For example, the $\nu =1/m$ 
FQHE ($m$ is an odd integer) corresponds to the relation $S=(\frac{m}2)(N-1)$.
The corresponding collective excitation gap determines the stability of the 
FQHE \cite{FQHE_book}. 

The specific properties of the surface LLs in TI films that modify the
inter-electron interaction strength within a single LL and change the 
stability of corresponding FQHE are, (i) the inter-LL coupling of LLs of 
two surfaces modifies the structure of the corresponding wavefunctions and 
changes the mixture of the $n$ and $n-1$ non-relativistic LL functions. 
This effect is visible only for the $n=1$ TI Landau level and is the strongest 
at small thickness of the TI film, for which the inter-LL mixture is strong 
(see Fig.~\ref{Topological0}), and (ii) the finite thickness of the TI film 
results in a finite width of the wavefunctions, which reduces the interaction 
strength and hence reduces the FQHE gaps.  

To illustrate the manifestation of these two effects, we present in 
Fig.~\ref{Topological2} the dependence of the $\nu =\frac13$ excitation gap 
on the thickness of the TI film for $n=0$ and $n=1$ LLs and a fixed magnetic 
field. For large thickness of the TI film (see Fig.~\ref{Topological2}(a)) the 
inter-LL coupling is weak and the surface LLs are decoupled. In this case the 
main effect of the TI film on the FQHE states is through the finite width of 
the LL wavefunctions, which results in a reduction of the FQHE gaps. The monotonic 
reduction of the gaps for both $n=0$ and $n=1$ TI Landau levels is clearly 
visible in Fig.\ref{Topological2}(a) for large $L_z$. The corresponding
gap in a thicker TI film is usually less than the gap in $n=0$ LL of graphene 
(shown by the red arrow in Fig.~\ref{Topological2}(a)). The reduction 
is more pronounced for the $n=1$ LL and for large $L^{}_z$ the gaps in the
 $n=1$ and $n=0$ LLs become comparable. Although the reduction of FQHE in 
a thick TI film is  relatively large, the FQHE should be observable with a 
larger gap at the $n=1$ surface LLs. 

\begin{figure}
\begin{center}\includegraphics[width=7cm]{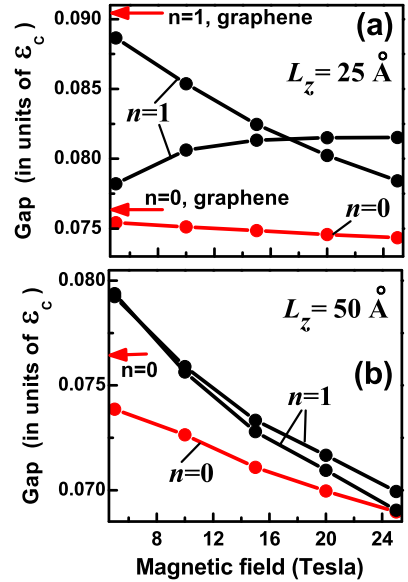}
\end{center}
\caption{\label{Topological1}
The $\nu=\frac13$ FQHE gap for different Landau levels of the
TI film: two $n=1$ LLs (black lines) and one $n=0$ LL (red line). 
The thickness of the film is (a) $L^{}_z=25$ \AA\ and (b) $L^{}_z=50$ \AA.
The red arrows show the FQHE gaps in the $n=0$ and $n=1$ Landau levels of 
graphene. The FQHE gaps were obtained numerically for a finite-size system of 
$N=9$ electrons and the parameter of sphere $S=12$. The energy is shown 
in units of the Coulomb energy, $\varepsilon^{}_C = e^2/\kappa\ell^{}_0$. 
}
\end{figure}

At a small thickness of the TI film, the inter-LL mixture becomes strong, which 
modifies considerably the behavior of the FQHE gaps in the $n=1$ LLs. Since the 
$n=0$ TI Landau level consists of the $n=0$ non-relativistic LL
wavefunction, the mixture of surface LLs has an weak effect on the interaction 
strength and the 
FQHE gaps in the $n=0$ LLs. As a result, the FQHE gap in the $n=0$ LL 
monotonically decreases with the TI film thickness within the whole range of
$L^{}_z$ due to the finite width of the LL wavefunctions. 

The $n=1$ TI Landau level shows a different behavior. Since the $n=1$ TI Landau 
level is a mixture of the $n=0$ and $n=1$ non-relativistic LL wavefunctions, the 
interaction potential becomes sensitive to the inter-LL coupling because such 
a coupling modifies the contribution of the $n=0$ and $n=1$ non-relativistic LL 
wavefunctions. As a result two $n=1$ TI Landau levels, although having almost 
the same energy, show different behavior as a function of $L^{}_z$. The difference 
between the two $n=1$ LLs is illustrated in Fig.~\ref{Topological2}(b), which shows 
the contributions of the $n=0$ and $n=1$ non-relativistic LL functions to the
TI LLs. Clearly, one of the $n=1$ TI LL (black lines) has a large contribution 
from the $n=0$ non-relativistic LL functions, while the other $n=1$ TI LL (blue 
line) has a large contribution from the $n=1$ non-relativistic LL functions. Due 
to these different contributions the FQHE gap of one of the $n=1$ LLs is large 
(almost equal to the gap of $n=1$ LL in graphene) for small $L^{}_z$ and decreases 
with increasing thickness of the TI film. The FQHE gap at the other $n=1$ LL is 
small at small $L^{}_z$ and with increasing $L^{}_z$ shows a well-pronounced 
maximum at $L^{}_z \sim 30$ \AA\ (see Fig.~\ref{Topological2}(a)). 

Another way to see these different properties of small and large thickness of 
the TI film is to study the dependence of the FQHE gaps on the magnetic field. 
In Fig.~\ref{Topological1} we present the dependence of the $\nu =\frac13$ 
FQHE gap on the magnetic field for two TI films with different thicknesses, 
$L^{}_z = 25 $ \AA\ (small), and $L^{}_z = 50$ \AA\ (large). 
For a large thickness of the TI film, $L^{}_z = 50$ \AA (see 
Fig.~\ref{Topological1}(b)) the surface LLs are decoupled. In this case the 
finite width of the LL wavefunctions results in a monotonic reduction of the 
FQHE gaps for both $n=0$ and $n=1$ LLs. The strength of such a reduction is 
characterized by the dimensionless width of the LL wavefunctions, expressed in 
units of the magnetic length, $\ell^{}_0$. With increasing magnetic field, the 
magnetic length decreases resulting in an increase of the dimensionless width of 
the LL wavefunctions. 

At a small thickness of the TI film, the inter-LL mixture is strong, which 
does not affect the FQHE gap in the $n=0$ LL but modifies the behavior of the FQHE 
gaps in the $n=1$ LL. For one of the $n=1$ LLs the FQHE gap decreases with increasing 
magnetic field, while for the other $n=1$ LL the FQHE gap increases 
with the magnetic field. Such an increase is the direct manifestation of 
inter-LL coupling. For both $n=1$ LLs the FQHE gap lies between the FQHE gaps 
of $n=0$ and $n=1$ LLs of graphene. 

Therefore, the FQHE can indeed be observed in the surface Landau levels of 
a topological insulator. The strength of the FQHE, which is characterized by 
the value of the excitation gap, is however reduced somewhat with increasing 
thickness of the TI film. This is due to an increase of the width in 
the $z$-direction of the surface Landau level wavefunctions. The finite 
thickness of the TIs notwithstanding, the FQHE gaps are the largest in the 
$n=1$ Landau level, which is similar to the case of a monolayer graphene 
\cite{Apalkov_06}. The reduction of the FQHE gaps is more pronounced in the 
$n=1$ Landau level. For the $n=1$ TI Landau levels the dependence of the 
FQHE gaps on the thickness of the film is strongly non-monotonic at a small 
thickness, which is due to strong inter-LL coupling. At a large enough thickness 
of the film, $L_z > 7$ nm, the gaps of the FQHE states in the $n=0$ and $n=1$ 
Landau levels become comparable. Experimental observation of these 
theoretical predictions, just as in the case of graphene 
\cite{Apalkov_06,Abanin_10,Ghahari_11}, would be an important advancement in 
understanding the nature of correlated Dirac fermions in this unique state of matter.

The work has been supported by the Canada Research Chairs Program of the 
Government of Canada.


\begin{thebibliography}{99}
\bibitem{note_1}
There are several ways the Dirac surface state of a TI differs from 
the Dirac cones in graphene. In the TI, a single Dirac cone can
form in the surface Brillouin zone, while in graphene they always come
in pairs. The Dirac cone of TI is robust against any time-reversal
invariant perturbations. Dirac cones in graphene have spin degeneracy
which is absent in the TI \protect\cite{Topo_LL_1}.
\end{thebibliography}
\end{document}